\documentclass{article}
\usepackage{dcase2022,amsmath,graphicx,times,booktabs,amssymb}
\usepackage[inline]{enumitem}
\usepackage[hyphens]{url}

\makeatletter
\def\bstctlcite{\@ifnextchar[{\@bstctlcite}{\@bstctlcite[@auxout]}}
\def\@bstctlcite[#1]#2{\@bsphack
  \@for\@citeb:=#2\do{%
    \edef\@citeb{\expandafter\@firstofone\@citeb}%
    \if@filesw\immediate\write\csname #1\endcsname{\string\citation{\@citeb}}\fi}%
  \@esphack}
\makeatother 

\title{Matching Text and Audio Embeddings: Exploring Transfer-learning Strategies for Language-based Audio Retrieval}

\name{Benno Weck$^{1,2}$,
      Miguel P\'{e}rez Fern\'{a}ndez$^{1,2}$, 
      Holger Kirchhoff\,$^{1}$,
      Xavier Serra$^{2}$
      }
\address{$^1$ Huawei Technologies, Munich Research Center, Germany\\
        \{firstname.lastname\}@huawei.com\\
        $^2$ Universitat Pompeu Fabra, Music Technology Group, Spain\\
        \{firstname.lastname\}01@estudiant.upf.edu, xavier.serra@upf.edu\\
 }

\begin{document}

\ninept
\maketitle
\bstctlcite{IEEEexample:BSTcontrol}

\begin{abstract}
We present an analysis of large-scale pretrained deep learning models used for cross-modal (text-to-audio) retrieval.
We use embeddings extracted by these models in a metric learning framework to connect matching pairs of audio and text.
Shallow neural networks map the embeddings to a common dimensionality.
Our system, which is an extension of our submission to the Language-based Audio Retrieval Task of the DCASE Challenge 2022, employs the RoBERTa foundation model as the text embedding extractor.
A pretrained PANNs model extracts the audio embeddings.
To improve the generalisation of our model, we investigate how pretraining with audio and associated noisy text collected from the online platform Freesound improves the performance of our method.
Furthermore, our ablation study reveals that the proper choice of the loss function and fine-tuning the pretrained models are essential in training a competitive retrieval system.
\end{abstract}


\section{Introduction}
\label{sec:intro}

The \emph{DCASE2022} challenge subtask 6b provides a platform to stimulate research in the underexplored problem domain of language-based audio retrieval \cite{xie_dcase_2022}.
The goal of this task is to find the closest matching audio recordings for a given text query.
A possible application for this task is a search engine for audio files in which a user can enter a free-form textual description to retrieve matching recordings.
Such systems need to draw a connection between the two modalities: audio and text.

Given the complex nature of both audio and text, we expect that a system can only perform well in this task if it can capitalise on a large amount of training data. 
Due to the novelty of the task, not many previous studies and systems exist for language-based audio retrieval and training data is still limited.
We instead turn to the fields of machine listening, specifically audio tagging, and natural language processing to draw inspiration from related problems and make use of existing resources such as pretrained models.
It has become a popular approach to use large-scale pretrained models in a transfer learning setup for tasks where only limited training data is available.

The goal of this work is to study a simple, generic cross-modal alignment system.
Our approach should be able to process audio and text independently to be used in a cross-modal retrieval context.
Therefore, we leverage the power of pretrained models and
a metric learning framework to semantically link the two modalities. 
We limit the complexity of our approach by employing the pretrained models with fixed weights and only train shallow network architectures to perform the alignment.
Additionally, this paper presents an analysis of our submission \cite{weck_aligning_2022} to the Language-based Audio Retrieval Task of the DCASE2022 Challenge.
With an ablation study, we investigate the impact of different training strategies on the performance of our system.
This helps us to understand the differences in performance between our system and other submissions to the challenge.

The remainder of this paper is structured as follows.
In the next section, we introduce the methodological framework of our system.
Section \ref{sec:experiments} explains the experiments that lead to our challenge submission and Section \ref{sec:results} presents the results of the submitted systems.
The results of additional experiments performed as an ablation study are discussed in Section \ref{sec:ablation}.
We summarise our findings in Section \ref{sec:Conclusion}.

\begin{figure}
    \centering
    \includegraphics[width=\columnwidth]{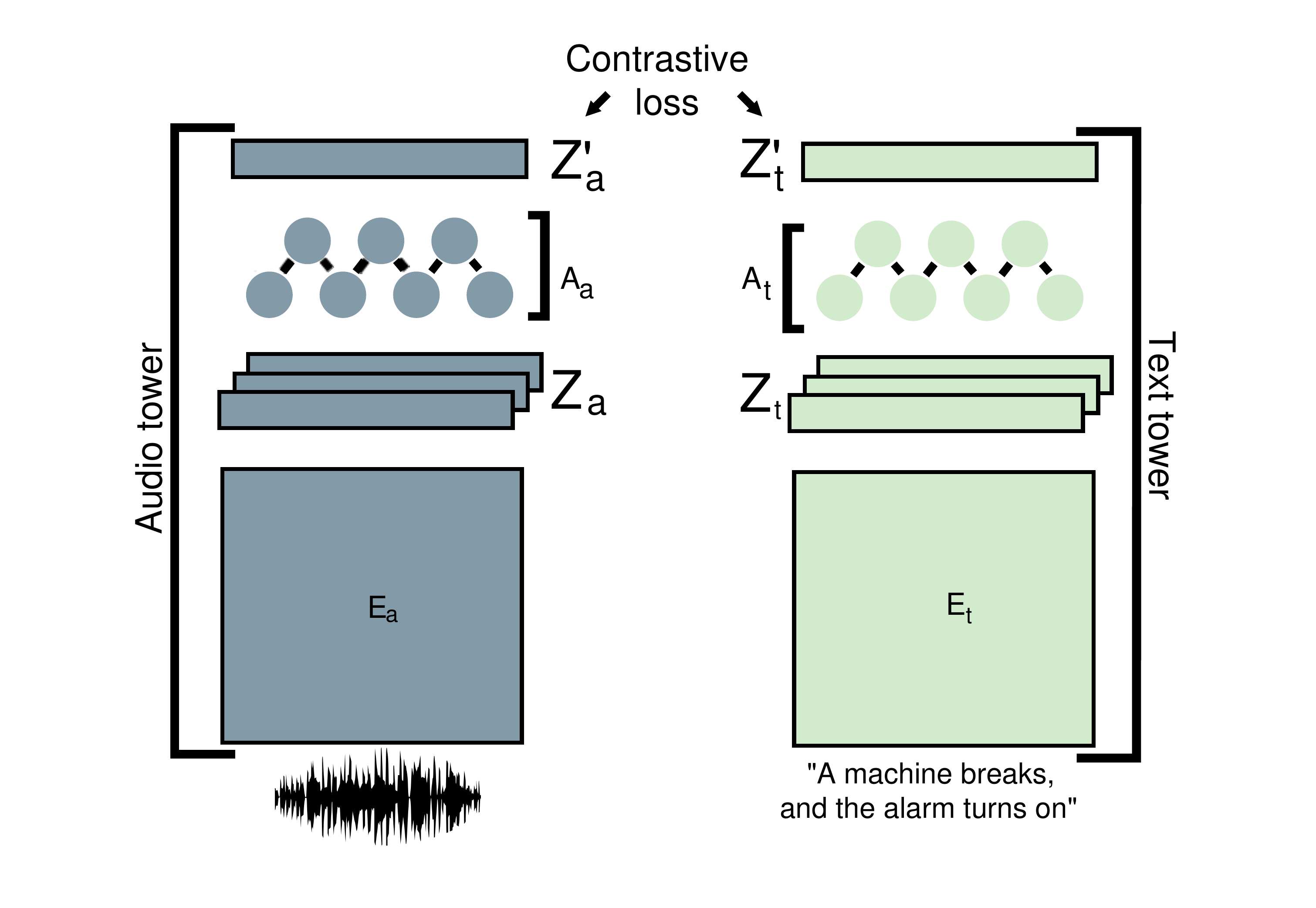}
    \caption{Overview of the architecture of our system. An audio tower and a text tower process the respective input data separately and produce a single embedding.}
    \label{fig:model}
\end{figure}

\section{Method}
\label{sec:method}
We adopt a metric learning \cite{mahmut_deep_2019} framework in our approach, which differs from a classification scenario used in related tasks such as audio tagging.
In a classification scenario, the outputs of a network are the predictions for the different classes and the features that characterise each of those classes remain in the intermediate layers of the network.
However, in metric learning, the goal is to obtain those features directly, so that the output of the network can be used to measure the similarity between two different inputs.
The features learned by the system can be referred to as an `embedding space'.
For each input, a network trained with metric learning will return an embedding $\mathbf{Z}\in\mathbb{R}^{F}$, where $\mathbb{F}$ is the size of the embedding, which is a hyper-parameter. 

Metric learning usually relies on `positive' and `negative' examples to teach the networks.
Positive examples are pairs of inputs that share some similarities, e.g., two sounds of birds singing.
Negative examples, on the other hand, contain dissimilar content, e.g., a recording of a bird singing and a car's ignition system.
The positive examples should be `closer' in the embedded space, while the negative ones should lie in different regions.
In our case, positive examples are audios and their corresponding descriptions.

Our system consists of two components -- an audio tower and a text tower -- to separately process the audio and text input.
Each tower is further divided into an encoder, $\text{E}(\cdot)$, and an embeddings' adapter, $\text{A}(\cdot)$.
As the audio encoder $\text{E}_a$ and the text encoder $\text{E}_t$, we employ pretrained models.
An overview of our method is presented in Figure~\ref{fig:model}. 

More specifically, an audio input $\mathbf{X}_a$ or a text input $\mathbf{X}_t$ are processed by $\text{E}_a$ and $\text{E}_t$, respectively, as
\begin{equation}
  \begin{aligned}
    \mathbf{Z}_a &= \text{E}_a(\mathbf{X}_a)\text{,}\\
    \mathbf{Z}_t &= \text{E}_t(\mathbf{X}_t)\text{,}
  \end{aligned}
\end{equation}
\noindent
where $\mathbf{Z}_i\in\mathbb{R}^{T_i \times F_i}, i \in \{a,t\}$ is a sequence of $T_i$ intermediate representations with $F_i$ features provided by the pretrained model (i.e., an embedding sequence).
Then, the adapters $\text{A}_a$ and $\text{A}_t$ will process $\mathbf{Z}_a$ and $\mathbf{Z}_t$ as
\begin{equation}\label{eqn:adapter}
  \begin{aligned}
    \mathbf{Z}'_a &= \text{A}_a(\mathbf{Z}_a)\text{,}\\
    \mathbf{Z}'_t &= \text{A}_t(\mathbf{Z}_t)\text{,}
  \end{aligned}
\end{equation}
\noindent
where $\mathbf{Z}'_{a},\mathbf{Z}'_{t}\in\mathbb{R}^{F'}$ are single embeddings and $F'$ denotes their dimensionality.
The intermediate embedding sequences $\mathbf{Z}_a$ and $\mathbf{Z}_t$ produced by the audio and text encoder respectively will differ in dimensionality. 
The main purpose of the adapters is to match the dimensionality of text and audio embeddings in order to enable comparisons.
We use the metric learning techniques described above to align the embedded spaces $\mathbf{Z}_a$ and $\mathbf{Z}_t$, so during training the adapters will learn to bring both into a common embedding space.

We experimented with two different losses.
The first is the contrastive loss \cite{chopra_learning_2005}, which we used for our submission to the DCASE2022 challenge.
Given the cosine similarity $s$ between a pair of embeddings with labels $l_1$ and $l_2$, the contrastive loss is defined by:
\begin{equation}
    L_{contrastive} =  
    \begin{cases}
    1 - s & \text{if}\ l_1 = l_2 \\
    \text{max}(0, s) & \text{otherwise.}
    \end{cases}
\end{equation}
The second loss that we use in our experiments is the Normalized Temperature-scaled Cross Entropy (NT-Xent) loss \cite{sohn_improved_2016}, which is used by the leading submissions in the DCASE2022 challenge.
For a more concise explanation of this loss, we refer the reader to the technical reports of the top-ranked teams \cite{xu_sjtu_2022,mei_language-based_2022}.

For the final application as a text-to-audio retrieval system, we compute the embedding of the text query $\mathbf{Z'}_t$ and compares it to all pre-computed embeddings $\mathbf{Z'}_a$ of the audio items in the dataset by means of the cosine similarity.
Ranking the audio items by their similarity score in descending order provides the retrieval results.
\section{Experiments}
\label{sec:experiments}

\subsection{Datasets}
As the main dataset in our work, we employ the development dataset provided for this challenge, \emph{Clotho v2} \cite{drossos_clotho_2020}, and use its official splits for training, validation, and final evaluation (testing).
We posit that the Clotho dataset is relatively small for the training of deep-learning-based retrieval systems and any system might benefit from additional training data.
Datasets combining audio and text are scarce, however, and the few that exist besides Clotho are either specific to a certain domain (e.g., urban soundscapes only \cite{martin-morato_diversity_2021}) or their audio content is not freely accessible \cite{kim_audiocaps_2019}.
This is why we decided to use weakly aligned text and audio pairs collected from the online platform Freesound \cite{font_freesound_2013}, which also served as the data source for Clotho.
Freesound allows users to upload an audio recording along with a textual description and a set of tags.
This type of metadata was used before to extend the training data of Clotho but in the context of an automated audio captioning task \cite{han_automated_2021}.
For simplicity and reproducibility, we limit ourselves to the \emph{dev} subset of the \emph{FSD50k} dataset \cite{fonseca_fsd50k_2022}.
We assume that the audios in this dataset closely resemble the challenge audio data as the dataset mainly comprises recordings of sound events.
Moreover, similarly to Clotho, audio clips are not longer than 30 seconds.
The descriptions and tags in the dataset contain rich information about the content of the audio clip as can be seen from the examples given in Table \ref{tab:fsd50k_examples}.
Nevertheless, the text data is noisy and also contains some undesired text.\footnote{For example: ``CAUTION: THIS PACK IS A CHEAP HOME RECORD. (But this one sounds a bit better)''}
To clean the descriptions we remove all HTML mark-up and limit each text to 500 characters in a pre-processing step.
To form a `sentence' out of the tags, we join them with a single white space in the order given in the dataset.
The \emph{dev} split of the FSDK50 dataset contains almost 44100 audio files and we use half of them.
By using descriptions and tag sequences, we can extend the training data by 40966 text-audio pairs (more than twice the amount of caption-audio pairs in the training subset of Clotho).
We refer to Clotho's data as `clean' and FSD50k's data as `noisy'.

\begin{table}
\centering
\footnotesize
\begin{tabular}{@{}p{0.48\columnwidth} p{0.47\columnwidth}@{}} 
\toprule
Description & Tags \\ 
\midrule
``Typing on a mechanical keyboard'' & ``click'',
      ``keyboard'',
      ``mechanical'',
      ``computer'',
      ``typing'',
      ``button''     \\
\addlinespace
  ``Pouring liquid in a shot glass, picking it up, drinking \& slamming it down (not too hard) on the table.''  & ``slam'',
      ``glass'',
      ``pour'',
      ``drink'',
      ``liquid'',
      ``alcohol'',
      ``shot'' \\
\addlinespace
   ``opening of shower curtain, turning shower on, water running, turning shower off, getting out'' & ``shower'',
      ``water'',
      ``bathroom'',
      ``bathtub'',
      ``human'' \\
\bottomrule
\end{tabular}
\caption{Hand-picked examples of descriptions and text labels from the metadata of the FSD50k dataset.}
\label{tab:fsd50k_examples}
\end{table}

\subsection{Evaluation \& Metrics}
We evaluate the ranked retrieval results generated by our systems with the same four metrics as the challenge organisers.
Specifically, we report three `recall at $k$' metrics (\emph{Recall@1},\emph{ Recall@5}, \emph{Recall@10}) and one `mean average precision at $k$' (\emph{mAP@10}), where a score for a given query is computed for the top-$k$ retrieved results and all scores are averaged over the entire set of queries.
We direct the reader to \cite{manning_introduction_2008} for an in-depth explanation of the metrics.

\begin{table*}[ht]
\centering
\begin{tabular}{@{}llllll@{}}
\toprule
 & \multicolumn{4}{c}{Development test set} & Challenge test set \\ 
\cmidrule(lr){2-5}\cmidrule(l){6-6}
 & Recall@1 & Recall@5 & Recall@10 & mAP@10 & mAP@10 \\ \midrule
Challenge baseline* & 0.03\phantom{0}
& 0.11\phantom{0}
& 0.19\phantom{0}
& 0.07\phantom{0}
& 0.061 \\
ensmbl\_5* \cite{xu_sjtu_2022} & 0.188 & 0.447 & 0.587 & 0.299 & 0.276 \\
Mei\_Surrey\_1* \cite{mei_language-based_2022} & 0.150 & 0.400 & 0.530 & 0.260 & 0.251\\
\addlinespace
ATAE & 0.071 (0.064 - 0.078) & 0.217 (0.206 - 0.228) & 0.325 (0.312 - 0.337) & 0.136 (0.128 - 0.143) & 0.114 \\
ATAE-ET & 0.064 (0.057 - 0.070) & 0.194 (0.184 - 0.205) & 0.288 (0.275 - 0.300) & 0.121 (0.114 - 0.128) & 0.113 \\
ATAE-EP-F & 0.067 (0.061 - 0.074) & 0.200 (0.189 - 0.210) & 0.299 (0.286 - 0.311) & 0.127 (0.120 - 0.134) & 0.121 \\
ATAE-NP-F & 0.072 (0.065 - 0.079) & 0.225 (0.214 - 0.236) & 0.325 (0.313 - 0.338) & 0.139 (0.131 - 0.146) & 0.128 \\
\bottomrule
\end{tabular}
\caption{Retrieval metrics for the four submitted systems, the two leading teams, and the challenge baseline.
The 95\% confidence intervals computed by jackknife resampling are given in parentheses.
Results marked with * were reported by the challenge organisers.
}
\label{tab:results}
\end{table*}

\subsection{Implementation details}
Our system is implemented by relying on the \emph{PyTorch} \cite{paszke_pytorch_2019} framework in connection with the \emph{pytorch-metric-learning} package \cite{musgrave_pytorch_2020}.
For the text processing, we employ the \emph{Transformers} library \cite{wolf_transformers_2020} and use the pretrained \emph{distilroberta-base} model as the text encoder.
This model is a compressed version of the original \emph{RoBERTa} model \cite{liu_roberta_2019} created by a knowledge distillation procedure \cite{sanh_distilbert_2019}.
It is smaller and faster than the original variant while retaining high performance on downstream tasks.
Similar to our previous work on audio captioning \cite{weck_evaluating_2021}, we decided to use the penultimate layer as the intermediate embeddings $\mathbf{Z}_t$.
The extracted text embeddings have a dimensionality $F_t$ of 768.

For the audio processing, we use a pretrained \emph{PANNs} model \cite{kong_panns_2020} as the audio encoder.
We follow the authors' suggestion and compute embeddings by taking the post-activation output of the penultimate layer of their \emph{CNN14} model.\footnote{Pretrained weights can be found at: \url{https://doi.org/10.5281/zenodo.3987831}}
All audio clips are resampled to a sampling rate of 32 kHz in a preprocessing step.
The extracted intermediate audio embeddings $\mathbf{Z}_a$ have a dimensionality $F_a$ of 2048.

We use simple feed-forward neural networks to adapt each embedding sequence to the common dimensionality. 
Both adapters consist of a two-layer perceptron with a layer size of 512 and a rectified linear unit (ReLU) as activation function after the first layer.
We use the average of all embeddings in a sequence as the final representation.

The system is optimised by minimising the contrastive loss with the Adam algorithm \cite{kingma_adam_2015} ($\alpha=0.001, \beta_{1}=0.9, \beta_{2}=0.999,$ and $\epsilon=10^{-8}$).
We do not fine-tune the encoder models in our approach and only optimise the adapters.
To form a minibatch we randomly select 32 audio-text pairs from the training set.
We compute the loss for every possible combination of similar and dissimilar samples (including text-to-text and audio-to-audio pairs) and take the mean across all non-zero loss values.
Every epoch the mAP@10 metric is computed on the validation dataset.
We start training with a learning rate of $0.0001$ and reduce it by a factor of 10 if no improvement was found for five epochs. 
Finally, the training is stopped after ten epochs with no improvement and the model weights are reverted to the checkpoint of the epoch with the highest score. 

\subsection{Submitted systems}
We submit four different configurations of our system.
All share the same model hyperparameter configurations but differ in the way the available training data was used to train them.
Specifically, we experiment with:
\begin{enumerate*}[itemjoin={{, }}, itemjoin*={{, and }}]
    \item adding no external dataset in our training
    \item extending the training data with noisy data from the FSD50k dataset
    \item pretraining with noisy and clean data and later fine-tuning with clean data only
    \item pretraining exclusively with noisy data and fine-tuning with clean data only.
\end{enumerate*}

In every training (also if we refer to it as pretraining or fine-tuning), we follow the optimisation procedure described above.
\paragraph*{ATAE: Aligned Text and Audio Embeddings}
In its standard configuration, our system is trained solely with the challenge development dataset Clotho.
We refer to it as `Aligned Text and Audio Embeddings' or \emph{ATAE} for short.

\paragraph*{ATAE-ET: Aligned Text and Audio Embeddings -- Extended dataset for Training}
Next, we want to investigate if adding extra training data helps to improve retrieval performance.
To achieve this we combine the noisy FSD50k and the clean Clotho data into a single training dataset.

\paragraph*{ATAE-EP-F: Aligned Text and Audio Embeddings -- Extended dataset for Pretraining -- Fine-tuning}
To balance out the potential negative effects of the noise in the training data, we fine-tune the trained ATAE-ET model by again training with the clean Clotho dataset.

\paragraph*{ATAE-NP-F: Aligned Text and Audio Embeddings -- Noisy dataset for Pretraining -- Fine-tuning}
Finally, to be able to better judge the effect of the noisy data for pretraining, we use the datasets in two separate training stages.
We first train a model on the noisy data and then fine-tune it on the clean dataset.

\section{Results}
\label{sec:results}

Table \ref{tab:results} compares the metrics achieved for our four systems with the challenge baseline and two of the leading submissions on the challenge development test set and the challenge test set. 
We follow the lead of the challenge organisers and report a jackknife approximated 95\% confidence interval for each metric \cite{mesaros_sound_2019}.
Based on the results on the development test set, we make the following observations.
First, our approach produces good quality results even in the standard training setup (ATAE: mAP@10 = 0.136 for the development test set).
Second, extending the challenge dataset with additional (noisy) training data significantly degrades retrieval performance (ATAE-ET: mAP@10 = 0.121).
Third, even fine-tuning the second system on the clean challenge dataset seems to give worse results (ATAE-EP-F: mAP@10 = 0.127) in comparison with simply training only with the challenge dataset (ATAE).
Fourth, our system first pretrained with noisy data only and then fine-tuned on the challenge dataset (ATAE-NP-F: mAP@10 = 0.139) improves on the performance of the first experiment but only slightly.
Finally, all of our submitted systems surpass the challenge baseline in each metric by a comfortable margin but are inferior to the best systems in the challenge.

Since the metrics of our best system (ATAE-NP-F) lie within the confidence intervals of our next best system (ATAE) and vice versa, we conclude that no significant difference is measurable between them.
These results suggest that no apparent advantage exists for our method in utilising additional noisy training data.
However, when comparing the two systems (ATAE \& ATAE-NP-F) on the challenge test set the advantage of pretraining with external data is more noticeable.
A possible explanation for this might be that the model pretrained with additional external data has better generalisation capabilities and is less affected by a shift in data distribution.

\section{Ablation study}
\label{sec:ablation}

Our approach is similar to the systems of the two top-ranked teams (\cite{xu_sjtu_2022,mei_language-based_2022}) in the DCASE2022 challenge, yet we fail to reach the same level of retrieval performance.
For example, analogous to us, both teams employ a two-tower architecture and shallow neural networks as adapter layers.
Their choice of pretrained models (e.g., PANNs \& RoBERTa) is also similar to ours.
The most striking differences between our and their submissions are that they decided to:
  \begin{enumerate*}[label=(\roman*), itemjoin={{, }}, itemjoin*={{, and }}]
  \item use NT-Xent as a loss function
  \item fine-tune the encoder models
  \item use the AudioCaps dataset \cite{kim_audiocaps_2019} in pretraining.
\end{enumerate*}
In view of this resemblance, we conduct additional experiments to investigate why a large gap in performance exists between our submission and the top-ranked systems.

We test five additive changes in training configuration.
The results for each of the configurations are computed from five training runs.
First, we employ the NT-Xent loss instead of the contrastive loss.
Second, we assess the impact of pair selection for the loss function on the retrieval metrics.
Our submission systems were trained considering not only text-audio pairs but also text-text and audio-audio pairs in the loss calculation.
Since samples from different training instances (i.e., with different labels) will be considered dissimilar but could contain semantically similar content (e.g., two different recordings of birds), this could harm the training process.
Therefore, we compare using only text-audio pairs in the loss calculation with using all possible pairs.
Third, we want to test if our approach is restricted by the fixed encoder models and can benefit if they are fine-tuned in the training process.
To limit the computational cost, we adopt the idea to only fine-tune the text encoder from a work in computer vision that showed that only fine-tuning the text model can help to train competitive text-to-image alignment models \cite{zhai_lit_2022}.
Fourth, we investigate the potential of pretraining with additional data.
As we saw from the results in Section \ref{sec:results}, pretraining with extra (noisy) data might help the model generalise better to unseen data.
Also, both leading teams adopt pretraining in their training process.
This is why we test if adding a pretraining stage relying on the entire \emph{dev} split of the FSDK50 dataset can enhance our system's performance.
Finally, we evaluate the benefits of fine-tuning both encoder models instead of only the text encoder similar to the approach in \cite{mei_language-based_2022}.

\begin{figure}[ht]
    \centering
    \includegraphics[width=\columnwidth]{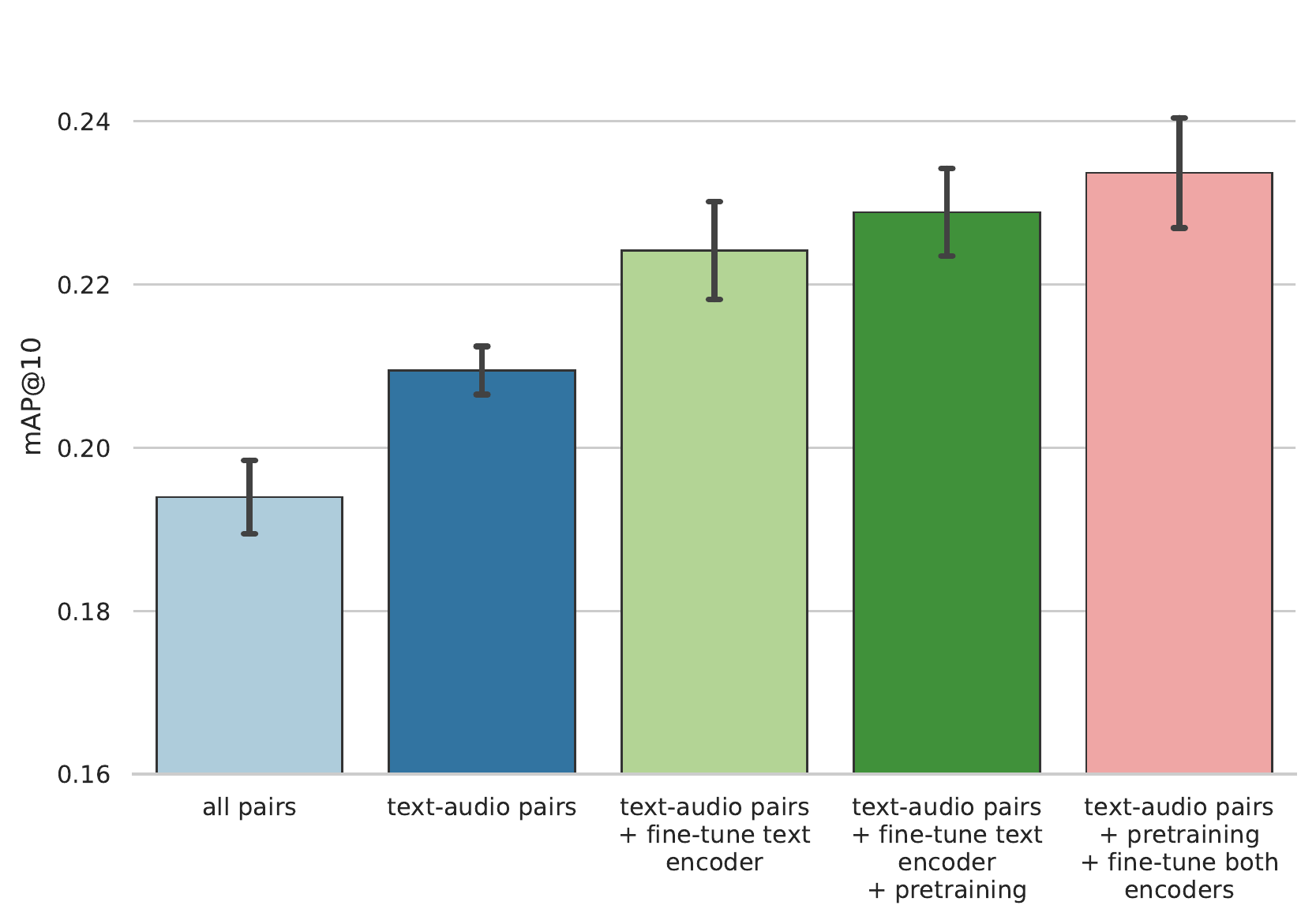}
    \caption{Comparison of the average retrieval results measured in mAP@10 on the development test set for different training configuration settings.
    The error bars show the standard deviation.
    }
    \label{fig:ablation1}
\end{figure}

Figure \ref{fig:ablation1} compares all ablation experiment configurations by the average mAP@10 achieved on the development test set.
What can be clearly seen in this figure is the accumulative increase in mAP@10 with every added change.
We find that replacing the contrastive loss with the NT-Xent loss (see `all pairs' in Fig. \ref{fig:ablation1}) already gives improved results in comparison with our challenge submission (mAP@10 = 0.193 compared to ATAE: mAP@10 = 0.136).
Only considering text-audio pairs in the NT-Xent loss, however, further improves the retrieval performance to mAP@10 = 0.209.
Furthermore, fine-tuning the text encoder model and including a pretraining stage adds to the improvement (mAP@10 = 0.224 and mAP@10 = 0.228, respectively).
As the last change, fine-tuning both encoder models results in the best score on average (mAP@10 = 0.233).
This comparison points to the conclusion that fine-tuning the encoder models and a pretraining stage are essential to achieve a high retrieval performance with our method.
However, with the small sample size, the results must be interpreted with caution as the difference between the last three settings might not be significant.

\section{Conclusion}
\label{sec:Conclusion}

We presented an analysis of our submission for the \emph{Language-based Audio Retrieval} subtask of the DCASE2022 challenge.
Our approach consists of extracting embeddings for the text and the audio through pretrained encoder models and mapping these embeddings to a shared space with a cross-modal alignment procedure.
The best system in our submission is a model that is first pretrained with noisy text-audio data collected from Freesound and later fine-tuned on the challenge dataset.
Even though our approach is similar to those of other teams we fall behind in the competition.
Through an ablation study, we show that a large part of the performance gap can be attributed to our choice of the loss function and the fact that we keep encoders fixed instead of fine-tuning them.
Moreover, we note promising results when pretraining our models with noisy data.
Future work should further investigate the use of large quantities of noisy data for pretraining.

\bibliographystyle{IEEEtran}
\bibliography{refs}

\end{document}